\documentclass[aps,prd,twocolumn,amsmath,superscriptaddress,amssymb,showpacs,floatfix,nofootinbib,longbibliography]{revtex4-1}
\pdfoutput=1
\usepackage{graphicx}
\usepackage{bm}
\usepackage{times}
\usepackage{slashed}
\usepackage{color}
\usepackage{aas_macros}
\usepackage{hyperref}
\hypersetup{
     colorlinks   = true,
     citecolor    = magenta,
     urlcolor     = magenta,
     linkcolor    = magenta
}

\usepackage{slashed}
\usepackage{lipsum}
\usepackage{subfigure}
\usepackage{multirow}
\usepackage{amsmath}
\usepackage{array} 
\usepackage{varwidth} 

\newcommand{\be}{\begin{equation}}
\newcommand{\ee}{\end{equation}}
\newcommand{\bea}{\begin{eqnarray}}
\newcommand{\eea}{\end{eqnarray}}

\begin{document}

\title{Self-Generated Cosmic-Ray Turbulence Can Explain the Morphology of TeV Halos}
\author{Payel Mukhopadhyay}
\email{payelmuk@stanford.edu; ORCID: orcid.org/0000-0002-3954-2005}
\affiliation{SLAC National Accelerator Laboratory, Stanford University, Stanford, CA 94039, USA}
\affiliation{Physics Department, Stanford University, Stanford, CA 94305, USA}

\author{Tim Linden}
\email{linden@fysik.su.se, ORCID: orcid.org/0000-0001-9888-0971}
\affiliation{Stockholm University and The Oskar Klein Centre for Cosmoparticle Physics,  Alba Nova, 10691 Stockholm, Sweden}

\begin{abstract}
Observations have shown that spatially extended ``TeV halos" are a common (and potentially generic) feature surrounding young and middle-aged pulsars. However, their morphology is not understood. They are larger than the ``compact" region where the stellar remnant dominates the properties of the interstellar medium, but smaller than expected in models of cosmic-ray diffusion through the standard interstellar medium. Several explanations have been proposed, but all have shortcomings. Here, we revisit a class of models where the cosmic-ray gradient produced by the central source induces a streaming stability that ``self-confines" the cosmic-ray population. We find that previous studies significantly underpredicted the degree of cosmic-ray confinement and show that corrected models can significantly inhibit cosmic-ray diffusion throughout the TeV halo, especially when similar contributions from the coincident supernova remnant are included. 
\end{abstract}

\maketitle

\section{Introduction}

TeV halos are a distinct class of galactic $\gamma$-ray emission sources characterized by their hard $\gamma$-ray spectrum, spatially extended and roughly spherically symmetric morphology, and coincidence with middle-aged pulsars. Building on initial observations by the High Altitude Water Cherenkov (HAWC) telescope of two TeV halos surrounding the Geminga and Monogem pulsars~\cite{Abeysekara:2017hyn, Abey:2017}, subsequent observations by HAWC, the High-Energy Steroscopic System (H.E.S.S) and Large High-Altitude Air Shower Observatory (LHAASO) have identified at least 8 TeV halo systems~\cite{2017PhRvD..96j3016L, 2017ATel10941....1R, 2018ATel12013....1B, 2019MNRAS.488.4074F, 2019PhRvD.100d3016S, Fang:2021qon, LHAASO:2021crt}. Dozens of other systems have been discussed as possible TeV halos, or TeV Halo/Pulsar Wind Nebulae (PWN) composite systems~\cite{2017PhRvD..96j3016L, Linden:2017blp, Profumo:2018fmz, 2019PhRvD.100d3016S} , a distinction which primarily depends on the definition used to distinguish standard PWNe from the more extended and diffuse TeV halos~\cite{2019PhRvD.100d3016S, Giacinti:2019nbu}. 

The size and morphology of TeV halos is unexpected. Early theoretical models by Aharonian and collaborators~\cite{Aharonian:1995gt, Aharonian:1995zz} predicted that energetic pulsars could produce an extended halo of TeV electrons. However, these studies utilized standard values for galactic diffusion, leading them to conclude that the resulting halos would be extremely large and have a surface brightness too dim to observe.\footnote{See also, later studies based on Milagro observations of Geminga, which noted source emission~\cite{Abdo:2007ad, 2009ApJ...700L.127A, Yuksel:2008rf}}. Current observations indicate that the morphology of TeV halos is diffusive -- matching predictions from~\cite{Aharonian:1995gt, Aharonian:1995zz} However, the diffusion coefficient that was calculated from these observations lies approximately two orders of magnitude below the standard value for galactic diffusion~\cite{Hooper:2017gtd, Abey:2017, Fang:2018qco, Profumo:2018fmz, DiMauro:2019hwn, DiMauro:2019yvh}. This puzzling feature is not explained, and produces halos with a high surface brightness and typical extension between 20--50~pc at TeV energies. 

A number of models have been posited to explain the relatively compact morphology of TeV halos -- all of which have significant drawbacks: The first assumes that TeV halos occupy regions of space with unusually small pre-existing diffusion coefficients~\cite{Abey:2017, Tang:2018wyr,Giacinti:2018yjc,Lopez-Coto:2017pbk} that are not necessarily correlated with the pulsar or associated supernova remnant. While such a model was reasonable when only Geminga and Monogem had detected halos -- it becomes less credible as more TeV halos are detected. The second utilizes anisotropic diffusion models which indicate that diffusion is primarily confined to one dimensional flux tubes on small distance scales that orient particle diffusion in the direction along the line-of-sight between the pulsar and Earth~\cite{Liu:2019zyj}. This restricts the spatial extension of $\gamma$-ray emission perpendicular to the source, but should produce highly asymmetric sources that are not observed in most cases. A third model exploits the transition between rectilinear propagation and diffusion to temporarily oriented high-energy cosmic-rays towards Earth so that they can begin cooling before diffusing \cite{2021:Recchia}. However, the efficiency of diffusion in these scenarios requires the pulsars be significantly more energetic than indicated by radio observations \cite{Bao:2021hey}. 

Finally, some models argue that the pulsar (or associated supernova explosion) generates turbulence that drives down the local diffusion coefficient through a streaming instability~
\cite{Evoli:2018aza, Fang:2019iym}. This model mirrors similar studies of inhibited diffusion near supernova remnants~\cite{DAngelo:2017rou}, which indicate that the net contribution of all supernova remnants may even dominate cosmic-ray diffusion throughout the Milky Way halo~\cite{Evoli:2018nmb}. In particular, the steep cosmic-ray gradient produced by a bright source generates Alfv\'{e}n waves that propagate outward along the cosmic-ray gradient. Once excited, these Alfv\'{e}n waves dominate the turbulence
spectrum at the scattering scale because they are naturally resonant with the injected cosmic rays. The idea of self-generated turbulence by cosmic ray streaming is not new one and is often taken into account to model cosmic ray acceleration in supernova remnant shocks and propagation in the galactic halo \cite{Skilling:1970,Lagage:1983,Roberto:2013}. 

Unfortunately, early models of cosmic ray propagation around pulsars with self-generated turbulence indicated that the pulsar is not energetic enough to inhibit diffusion on such large scales~\cite{Evoli:2018aza, Fang:2019iym}. Additional contributions from the coincident supernova could potentially boost this effect. However, mature pulsars such as Geminga have often moved far from their parent SNR making it difficult to explain why the TeV emission then remains centered on the quickly moving pulsar~\cite{Fang:2019iym}.

In this paper we revisit models of cosmic-ray self-confinement. We first correct an error that affected previous results from Ref.~\cite{Evoli:2018aza}. In our corrected models, both the amplitude and duration of cosmic-ray self-confinement is significantly increased, making pulsars more than capable of powering cosmic-ray self-confinement in 1D simulations. We show that these models remain robust even when: (1) additional turbulence damping terms are included, (2) the pulsar spectrum is softened, or (3) the radius of the 1D flux tube is increased. In simulations where the initial diffusion coefficient is unsuppressed in three dimensions, the effect of cosmic-ray self-confinement decreases too rapidly with increasing radius. However, models that include additional contributions from the coincident SNR confine cosmic-rays on 10~pc scales. While fully consistent models of cosmic-ray self-confinement lie beyond the scope of this work, our models indicate that pulsars are energetically capable of confining their own cosmic-ray populations on Myr timescales.

\section{Model of the pulsar source}
\label{pulsar source}

We model the pulsar as a point-source that begins injecting electrons at time $t = 0$. Throughout this paper we use `electrons' to refer to $e^+e^-$ pairs unless otherwise specified. The spindown power ($L(t)$) of the pulsar is assumed to be:

\begin{equation}
    L(t) = L (t = 0) (1 + t / \tau)^{-2}
\end{equation}

\noindent where $L(t = 0)$ is the spindown power at $t = 0$, $t$ is the time and $\tau$ is the spindown timescale of the pulsar. The pulsar source function for 1D or 3D spherically symmetric propagation of cosmic rays is taken to be, respectively:

\begin{eqnarray}
\label{source_1D}
    Q^{1D}_e(z,p,t) = Q_0(t) \left(\frac{p}{m_e}\right)^{-\eta} e^{-\frac{p}{p_m}} \frac{\exp({- z^2 / 2 \sigma^2})} {\pi R^2 \sqrt{2 \pi \sigma^2}}  \nonumber \\
     \\
Q^{3D}_{e}(r,p,t) = Q_0(t) \left(\frac{p}{m_e}\right)^{-\eta} e^{-\frac{p}{p_m}} \frac{\exp({- r^2 / 2 \sigma^2})} {(2 \pi \sigma^2)^{3/2}} \nonumber
\end{eqnarray}

\noindent where $m_e$ is the electron mass, $\sigma$ is the source size, $\eta$ is the spectral index, $p_m$ is the momentum cutoff beyond which the source injection exponentially falls, $R$ is the flux tube radius for the case of 1D diffusion and $Q_0$ is a normalization constant. For the reference case considered in this paper, we set $\sigma = $ 1 pc in both models and $R = 1$ pc in the 1D case. 

The normalization $Q_0$ is determined such that the total power in electrons integrated over momenta is equal to the simultaneous spindown power of the pulsar. In reality, this term is modified by some efficiency factor that determines the fraction of the spindown power that is channeled to electrons. Observations of TeV halo luminosities indicate that both young and middle-aged pulsars convert a sizable fraction, $\mathcal{O}$(9--30\%), of their spindown power into electrons \cite{Hooper:2017gtd,Profumo:2018fmz,Johannesson:2019jlk,Evoli:2020szd}. For our reference case of the Geminga, we take $L (340 ~ \mathrm{kyr}) = 3.7 \times 10^{34}$ erg/s \cite{Manchester:2005} setting the age to 340 kyr and $\tau = 10$ kyr. We set the efficiency parameter to $\alpha$~= 0.1, corresponding to a 10\% electron injection efficiency.  

\section{Cosmic Ray Transport Modeling}
\label{sec:1D_transport}

The coherence length of the galactic magnetic field is expected to be somewhere between $L_c \sim$ 1--100 pc \cite{Iacobelli:2013fqa,Haverkorn:2008tb,Lopez-Coto:2017pbk}. It likely varies substantially throughout the galaxy, with shorter coherence lengths in regions with higher stellar and gas densities. Cosmic ray diffusion on distance scales smaller than the coherence length can be approximated by 1D diffusion in a flux tube. On larger scales, it is more appropriate to utilize 3D diffusion models. In this paper, we separately consider cosmic ray transport for both 1D and 3D geometries, ignoring the transition from 1D to 3D propagation for diffusion distances near $L_c$. In this section, we first detail a 1D treatment of diffusion, and then subsequently discuss several adjustments needed to consider particle propagation in 3D.

\subsection{Diffusion Equation in 1D}
Particle diffusion in a cylindrically symmetric 1D flux tube can be described by the following transport equation: 

\begin{eqnarray}
\label{eq:transport}
    \frac{\partial f}{ \partial t} = Q_e(p,z,t) - u_A \frac{\partial f}{\partial z} + \frac{\partial}{\partial z} \left( D(p,z,t) \frac{\partial f}{\partial z}\right) \nonumber \\
    + \frac{\partial u}{\partial z} \frac{p}{3} \frac{\partial f}{\partial p} - \frac{1}{p^2} \frac{\partial}{\partial p} \left( p^2 \frac{dp}{dt} f \right)
\end{eqnarray}

\noindent where $f$ is the phase space distribution function, $u_A$ is the advection speed, which we assume to be the Alfv\'{e}n velocity given by \mbox{$v_A = B_0 / \sqrt{4 \pi m_i n_i}$} ($B_0$ is the background magnetic field, $m_i$, $n_i$ represent the mass and number density of ions in the plasma), and $D(p,z,t)$ diffusion coefficient, which varies in momentum, space and time.

\subsection{Energy Loss Terms}
The last term, which includes $\frac{dp}{dt}$, takes into account the energy losses of $e^+e^-$ pairs. The energy loss rate is due to a combination of synchrotron radiation and inverse Compton scattering \cite{Blumenthal:1970gc} and is given by:

\begin{eqnarray}
\label{eq:energy_loss}
    - \frac{dp}{dt} = \sum_i \frac{4}{3} \sigma_T \rho_i S_i(E) \left(\frac{E_e}{m_e}\right)^2 + \frac{4}{3} \sigma_T \rho_{mag} \left(\frac{E_e}{m_e} \right)^2
\end{eqnarray}

\noindent where $\sigma_T$ is the Thomson cross section, $\rho_{mag}$ is the magnetic field energy density and the sum over in the expression with $\rho_i$ is carried out over the various radiation backgrounds consisting of the cosmic microwave background (CMB), infrared radiation (IR), ultraviolet emission (UV) and starlight. We choose $\rho_{CMB} = 0.260$ eV/cm$^3$, \mbox{$\rho_{IR}$ = 0.6 eV/cm$^3$}, \mbox{$\rho_{star} = 0.6$ eV/cm$^3$} and $\rho_{mag}$ = 0.025 eV/cm$^3$ for a background magnetic field of $B_0$ = 1 $\mu G$ \cite{Hooper:2017gtd}. At very high energies, inverse Compton scattering is further suppressed by Klein-Nishina effects (encoded in $S_i(E)$) which have also been taken into account following the analytic formalism of \cite{Schlickeiser:2009qq}, the accuracy of which has recently proven insufficient for the 1\% measurements of electron spectra by AMS-02~\cite{2021:FangK, DiMauro:2020cbn}, but is more than sufficient for our purposes here.  

\subsection{Self-Generated Turbulence and Non-Linear Damping}
\label{subsec:self-generated-diffusion}

The confinement of cosmic rays in our model depends on the generation of magnetic turbulence by the sharp gradient produced in the cosmic-ray population. Specifically, the diffusion coefficient ($D$) is related to the spectral power $W$ as given in Ref.\cite{Skilling:1971}

\begin{equation}
\label{eq:diffusion}
    D(p,z,t) = \frac{4}{3 \pi} \frac{r_L(p) v(p)}{k_{res} W_{res}} 
\end{equation}

\noindent where $r_L(p)$ is the Larmor radius and $v(p)$ is the particle speed, which we assume to be equal to the speed of light $c$. $k_{res} = 1/r_L(p)$ is the resonant wavenumber and $W_{res}$ is the spectral power computed at the resonant wavenumber. The evolution of $W$ can then be calculated by: 

\begin{equation}
\label{eq:wave}
    \frac{\partial W}{\partial t} + v_A \frac{\partial W}{\partial z}= (\Gamma_{CR} - \Gamma_D) W
\end{equation}

\noindent where the wave growth rate $\Gamma_{CR}$ for 1D propagation is given by:

\begin{equation}
\label{eq:growth}
    \Gamma_{CR} = \frac{16 \pi^2}{3 B_0^2} \frac{v_A}{k W(k)} \left( p^4 v(p) \frac{ \partial f}{\partial z}\right)_{p_{res}}
\end{equation}

\noindent with $\Gamma_{CR}$ being the growth rate of Alfv\'{e}n waves due to the resonant streaming instability \cite{Skilling:1971}. 

The turbulence will also be damped by a combination of several different processes. Our default models focus on non-linear wave damping (NLD) terms, which have been found to be the dominant damping term in regimes with similar magnetic field strengths and gas densities as observed near Geminga~\cite{Evoli:2018aza}. However, in the next section we will also discuss the impact of additional damping terms. For the NLD term, we follow the prescription of \cite{2003:Ptuskin} and set:

\begin{eqnarray}
\label{eq:damping}
\begin{split}
    \Gamma_{NLD} &= (2 c_K)^{-3/2} |v_A| k^{3/2} W^{1/2} \;\; (\mathrm{Kolmogorov})  \\
    \\
    &= (2 c_K)^{-3/2} |v_A| k^{2} W \;\;\;\;\;\;\;\;\;\; (\mathrm{Kraichnan}) 
\end{split}
\end{eqnarray}

\noindent where $c_K = 3.6$. Eq.(\ref{eq:damping}) is equivalent to having a Kolmogorov or Kraichnan description of the cascade given by $\frac{\partial}{\partial k} \left( D_{kk} \frac{\partial W}{\partial k}\right)$ in the limit where the cascade damping timescale is faster than any other damping timescales \cite{2004:Brunetti}. We note that it was this term in Ref.~\cite{Evoli:2018aza} which included a numerical error, where c$_K$ was accidentally set to be 0.21 (though the value was correctly listed in the paper). The error in this value significantly increased (by more than an order of magnitude) the rate of non-linear damping, significantly decreasing both the duration and the maximum inhibition of cosmic-ray diffusion throughout that work. 

\subsection{Additional Damping Processes}
In our default models, we consider only the NLD processes for turbulent damping. However, several additional dynamic processes can counteract the generation of magnetic turbulence and relax the diffusion coefficient to standard ISM values. Most notable is the ion neutral damping (IND) mechanism which occurs in partially ionized plasmas \cite{,1956:Piddington,1969:Kulsrud}. The IND mechanism dissipates wave energy through a viscosity term produced by charge-exchange interactions between ions and neutral atoms which cause neutral particles to begin participating in hydromagnetic phenomena. 

The damping rate depends on how the wave frequency $\omega_k = v_A k$ compares to the ion-neutral collisional frequency $\nu_{in} = 8.4 \times 10^{-9} \times (n_n / cm^{-3}) \times (T / 10^4 ~K)^{0.4}$ s$^{-1}$, where $n_n$ is the neutral atom number density and $T$ is the plasma temperature. The IND damping rate also depends on the ratio of ion to neutral densities $f_i = n_i / n_n$. In the limit $f_i >> 1$, which best describes interstellar regions that lie outside of dense molecular clouds, the IND damping rate is well approximated by Ref.~\cite{1982:Zweibel}:

\begin{equation}
    \Gamma_{IND} \simeq \frac{\nu_{in}}{2} \frac{\omega_k^2}{\omega_k^2 + f_i^2 \nu_{in}^2}
\label{eq:gamma_IND}
\end{equation}

For our default model, we assume $n_i = 1$ cm$^{-3}$ and  \mbox{$n_n \sim 10^{-2}$ cm$^{-3}$}, representing a neutral fraction of 1\%. These values are well-motivated by observations indicating the absence of H$\alpha$ emission from Geminga's bow shock nebula, which indicate that the neutral fraction surrounding the pulsar is less than 1\% \cite{Caraveo:2003}. In this case, we find that the IND damping timescale at 10 TeV is $\frac{1}{\Gamma_{IND}} > 10^6$ kyr, a value that significantly exceeds the NLD damping timescale of $ \frac{1}{\Gamma_{NLD}} \sim O(100 ~\mathrm{kyr})$. Based on this line of reasoning, the IND term was ignored in Ref.~\cite{Evoli:2018aza} and is similarly ignored in our default analysis. 

However, we note that the IND term can be significantly more important at low energies, or in regions with higher neutral gas densities. In order to determine the diffusion coefficient over the full energy range, we thus include IND damping for some simulations using neutral fractions of 1\% and 10\%, finding that such models significantly increase the diffusion coefficient at low energies. 

In addition to the NLD and IND damping mechanisms, additional damping terms such as Nonlinear Landau damping (NLLD) and linear turbulent damping (also known as the Farmer-Goldreich: FG damping), have been considered in the literature, but are not included in this analysis. NLLD damping effectively converts wave energy into heat \cite{1978:Kulsrud,2013:Weiner}. If this damping term dominates, the turbulent cascade is suppressed and the NLD cascade damping does not develop. However, the effectiveness of this damping term in a turbulent plasma is not clear. The functional form for the damping \cite{1978:Kulsrud,2013:Weiner} is obtained from a simplified system where two Alfv\'{e}n waves (with slightly different wavelengths) travel in opposite directions, forming beat waves that interact with the thermal ions and absorb the wave energy. A critical assessment of the role of this mechanism, using for example, Particle-in-cell (PIC) simulations, is still missing \cite{2019:Blasi}. Additionally, we note that if NNLD damping were dominant through the Galactic volume, then the turbulence injected by Supernova remnants would be inefficient to explain the diffusion coefficient observed in the Galaxy \cite{2003:Ptuskin}. 

The turbulent or FG damping term is given by \mbox{$\Gamma_{FG} \sim \frac{v_A}{\sqrt{L_{inj} r_L}}$}, where $v_{A}$ is the Alfv\'{e}n speed and $L_{inj}$ is the injection scale of the turbulence \cite{Farmer:2003mz}. For reasonable magnetic field strengths around Geminga, $B \sim 1$ $\mu G$ \cite{2019:LiuR}, the timescale associated to this damping term is $\sim 500$ kyr for $p \sim 10$ TeV particles, which is generally longer than the damping timescales associated to the NLD damping. Therefore, for the models considered in this paper, FG damping is expected to be subdominant. We note, however, that FG damping may become relevant for TeV halos formed in regions with larger interstellar magnetic fields.

\subsection{Adjustments for Diffusion in 3D}
As noted earlier, for cosmic ray diffusion on distance scales larger than the galactic coherence length, the diffusion is expected to be better represented by particle propagation in 3D -- although diffusion can be more complicated in the transition region where TeV halos may reside. In this subsection, we consider several adjustments to our model which are necessary to consistently model diffusion in a regime where particle propagation in all three spatial dimensions is initially unsuppressed.  We additionally note that TeV halo observations indicate that the halo is roughly spherical (or at least, not a linear tube), which strengthens the case for considering 3D diffusion models. We first note that in Eq.~\ref{source_1D}, we have provided the pulsar morphology in both the 1D and 3D spherically symmetric case, but find that these differences do not significantly impact the results of our study. 

More importantly, spherically symmetric particle diffusion in 3D space is governed by the 3D diffusion equation, given by:

\begin{equation}
\begin{aligned}
\label{eq:transport3D}
    \frac{\partial f}{ \partial t} = Q^{3D}_e(p,r,t) - v_A \frac{\partial f}{\partial r} + \frac{1}{r^2}\frac{\partial}{\partial r} \left( r^2 D(p,r,t) \frac{\partial f}{\partial r}\right) 
    &\\ + \frac{2 u_A}{r} \frac{p}{3} \frac{\partial f}{\partial p} 
    - \frac{1}{p^2} \frac{\partial}{\partial p} \left( p^2 \frac{dp}{dt} f \right)
\end{aligned}
\end{equation}

Finally, in addition to changes in the diffusion term, the wave growth rate in a spherically symmetric 3D geometry is given by:

\begin{equation}
\label{eq:growth_3D}
    \Gamma^{3D}_{CR} = \frac{16 \pi^2}{3 B_0^2} \frac{v_A}{k W(k)} \left( p^4 v(p) \frac{ \partial f}{\partial r}\right)_{p_{res}}
\end{equation}

where essentially the $\frac{\partial f}{\partial z}$ in eq. \ref{eq:growth} is replaced by the gradient in 3D spherical geometry \cite{2008:Ptuskin}. 

\subsection{Contributions from a Supernova Remnant}

Finally, we study a scenario where, in addition to electrons injected by the pulsar, a supernova remnant (SNR) also injects protons into the interstellar medium (see~\cite{Fang:2019iym}). In this case, the self-generation of waves is governed by the combined distribution function of both protons and electrons. The SNR source term is taken to be: 

\begin{multline}
Q^{SNR}(r,p,t) = Q_0(t) \left( \frac{p}{m_p}\right)^{-\eta_{SNR}} e^{-\frac{p}{p_{c}}} \\ 
\times \frac{\exp{(-r^2 / 2\sigma_{SNR}^2})}{(2 \pi \sigma_{SNR}^2)^{3/2}}
\label{eq:SNR_injec}
\end{multline}

\noindent where $Q_0$ is given by:

\begin{equation}
    Q_0(t) = \frac{\xi_{CR} E_{SN}}  {T_{SN} I}
\label{eq:Q_0}
\end{equation}

\noindent where $T_{SN}$ is the time over which the SNR provides injection power. For this paper, we assume $T_{SN} = 1000$ yr with the SNR injection stopping after this time. $I$ is a constant obtained by equating the total energy in protons to the fraction of SNR kinetic energy transmitted to protons. $\xi_{CR}$ is the fraction of the supernova kinetic energy transmitted to cosmic rays. 

Protons and electrons are treated separately and for each component, the propagation equation (Eq.~\ref{eq:transport3D}) is solved. Protons are assumed to have zero energy losses while electrons are taken to lose energy according to Eq. \ref{eq:energy_loss}.

\section{Numerical Modeling}
\label{sec:numerics}

The diffusion of particles and growth/damping of turbulence represent a set of coupled differential equations which we solve using an explicit finite-difference scheme. The timestep $\Delta t$ and spatial gridstep $\Delta x$ are chosen such that the system is stable. In this paper we choose $\Delta t$ = 0.05 yr and $\Delta x$ = 1 pc. The momentum grid has 120 logarithmic gridpoints between $p_{min} = 1$ GeV and $p_{max} = 10^6$ GeV. The results have been confirmed for a range of spatial and momentum grids. The validity of the results has also been tested in an unconditionally stable semi-explicit Crank Nicolson scheme.  

\begin{figure}
\centering
\includegraphics[width=.48\textwidth]{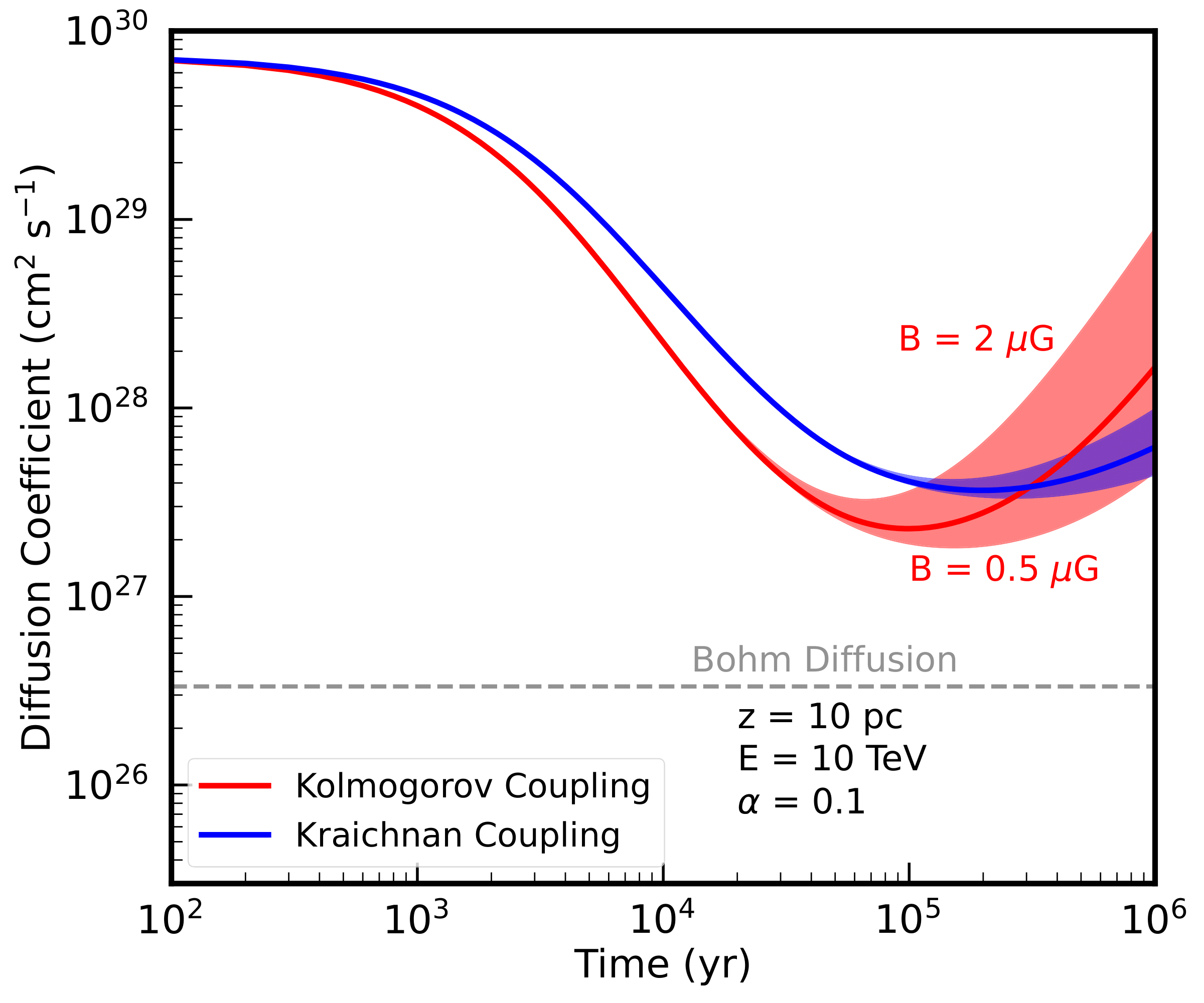}
\caption{The diffusion coefficient at a radius of 10 pc and an energy of 10 TeV as a function of the pulsar age in the 1D case. Two example scenarios, of Kolmogorov and Kraichnan turbulence, are shown. The shaded bands show results for magnetic field strengths ranging from 0.5--2 $\mu G$. In both cases, a strong suppression of the diffusion coefficient ($\sim 2-3$ orders of magnitude) is observed. The suppression is seen to persist over the Myr timescale in our simulation.}
\label{fig:kol_kraich}
\end{figure}

For our 1D fiducial model, we use $\alpha$ = 0.1, $B_0=1$ $\mu G$, $\sigma$ = 1 pc and set the background turbulence to have a Kolmogorov spectrum with $D = 3.466 \times 10^{28} ~ p_{\mathrm{GeV}}^{1/3}$ cm$^2$ s$^{-1}$. We take the background ion number density $n_i = 1 ~ \mathrm{cm^{-3}}$. We set the momentum dependence of the electron injection spectrum to be $\propto p^{-3.5} \exp(-\frac{p}{100 ~\mathrm{TeV}})$ over a range between 1~GeV and 10$^{7}$~GeV. The flux tube radius, $R$ is assumed to be 1 pc. The outer boundary of the flux tube is taken to be $z_b = 500$ pc, an unrealistically large number chosen such that boundary conditions do not affect our result near the pulsar. We set the electron distribution, $f$, to be 0 at the outer boundaries of the flux tube. The initial condition is set to be $f(z,p) = 0$ at $t = 0$.

For the 3D fiducial model, we keep most parameters to be the same as the fiducial 1D model, but set the default pulsar efficiency to be $\alpha$ = 1.0. The spatial gridstep is set to be $\Delta r = 1$ pc. The 3D boundary condition is set so that the electron distribution $f$ is zero at the outer boundary of the spherical volume considered, at $r_b = 500$ pc. The inner boundary of the simulation is set at $r_{in} = 0.1$ pc, instead of $r = 0$ to avoid singularities. For the inner boundary condition, we set $f(r_0,p) = f(r_1,p)$, where $r_0$ and $r_1$ denote the first two points of the radial grid. 

For 3D models that include an SNR, we set the SNR kinetic energy to be $E_{SN} = 10^{51} $erg, the SNR size, $\sigma_{SNR} = 1$ pc, SNR spectral index, $\eta_{SNR} = 4.2$ and fraction of energy transmitted to protons, $\xi_{CR} = 0.2$. The electron distribution function ($f_e$) and proton distribution function ($f_p$) are evolved separately with identical boundary conditions in each case.

\section{Results}
\subsection{1D model}
\label{sec:result_1D}

\begin{figure}[t]
\centering
\includegraphics[width=.48\textwidth]{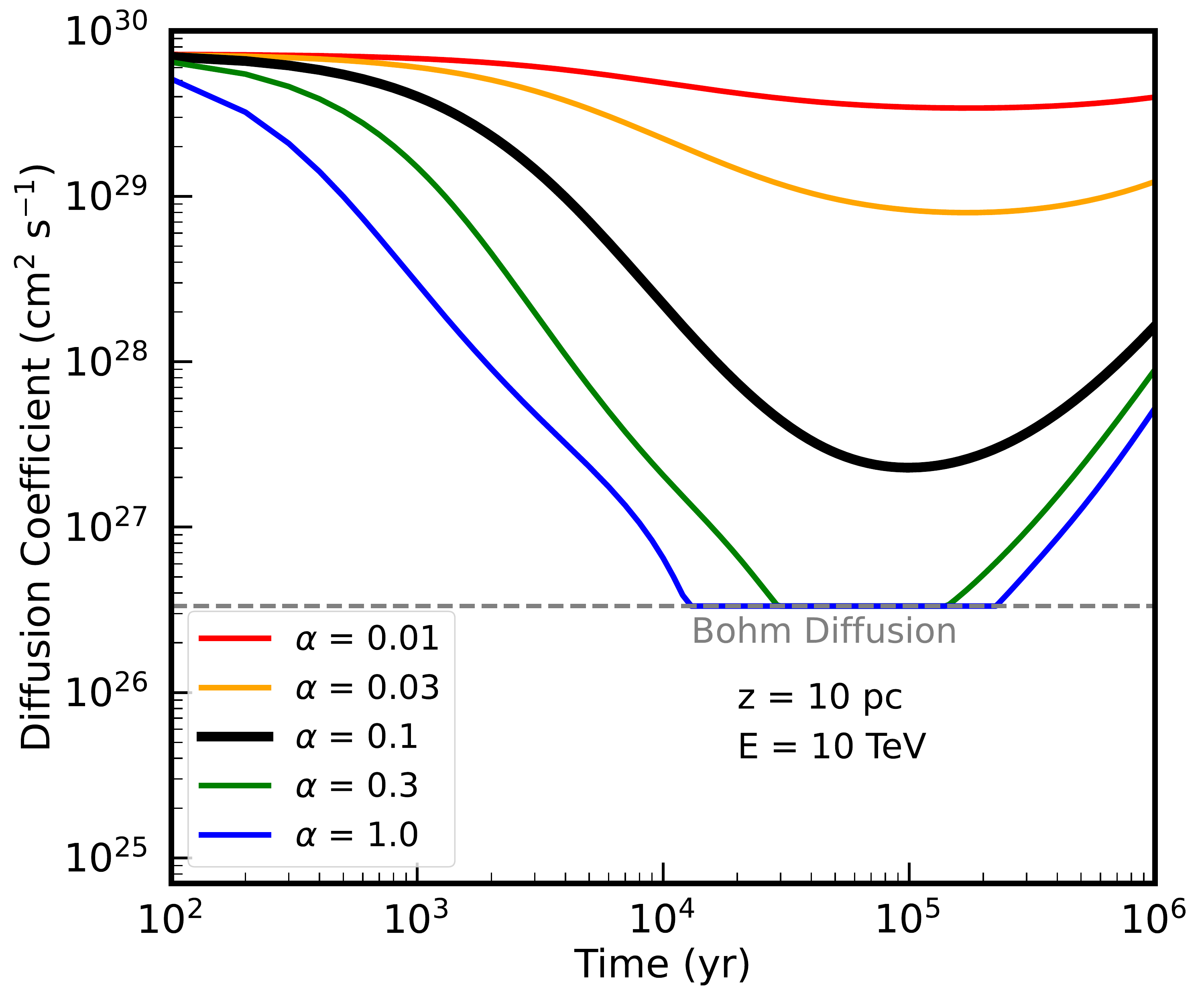}
\caption{The inhibition of the diffusion coefficient at an energy of 10~TeV and a distance of 10~pc from the location of the pulsar as a function of time. Curves correspond to different efficiencies in the conversion of the pulsar kinetic energy into e$^+$e$^-$ acceleration. Models with high efficiencies produce diffusion coefficients that fall below the Bohm limit, at which point our diffusion model may no longer be valid. In our models, we do not allow the diffusion coefficient to drop below the Bohm limit at any gridpoint.}
\label{fig:eff_dep}
\end{figure}

In Fig.~\ref{fig:kol_kraich}, we show the time evolution of of the diffusion coefficient for models utilizing the Kolmogorov and Kraichnan wave-damping phenomenology in our 1D simulation. We note that the initial turbulence spectrum is set to the Kolmogorov values in each case. We show results for values of the magnetic field strength ranging between 0.5--2 $\mu G$. The injection efficiency assumed for this plot $\alpha = 0.1$. The result is plotted at an electron energy of 10~TeV and at a distance 10~pc from the pulsar, roughly corresponding to the energy and distance scales corresponding to HAWC observations of the Geminga TeV halo~\cite{Abeysekara:2017hyn}. For both the Kolmogorov and Kraichnian cases, the diffusion coefficient decreases by more than 2.5 orders of magnitude compared to the background ISM value, reaching minimum values between 100--300~kyr, consistent with the ages of the Geminga and Monogem halos. Notably, the suppressed diffusion coefficient persists throughout the 1~Myr timescale of our simulation.

The suppression of the diffusion coefficient is governed by the growth of Alfv\'{e}n waves governed by $\Gamma_{CR}$, which induces a period of highly inhibited diffusion. The subsequent increase in the diffusion coefficient at late times is governed by the relaxation term $\Gamma_D$. Our models indicate that the diffusion coefficient does not return back to ISM values within the 1~Myr timescales of our simulation for either the Kolmogorov of Kraichnan models, with Kraichnan models having an even longer relaxation timescale~\cite{Evoli:2018aza}.

Our results are markedly different from those obtained in Ref.~\cite{Evoli:2018aza} (See Fig.~1 of Ref.~\cite{Evoli:2018aza} for a direct comparison.) While both studies found that cosmic-ray self-confinement could inhibit diffusion throughout the halo by 2.5--3 orders of magnitude at 10 TeV, our study requires only a 10\% conversion of the pulsar spindown power into electron pairs to achieve this effect, while Ref.~\cite{Evoli:2018aza} required a 100\% conversion of the pulsar power. More importantly, our models indicate that the diffusion coefficient requires more than 1~Myr to rebound to ISM values, while the relaxation time in Ref.~\cite{Evoli:2018aza} was much shorter and the diffusion coefficient returned to ISM values within 100~kyr. We have confirmed with the authors that the discrepancy between these results is due to an error in the relaxation term computed in Ref.~\cite{Evoli:2018aza}. The results presented here have been checked against a corrected version of the code used in \cite{Evoli:2018aza}, finding good agreement.

\begin{figure}[t]
\centering
\includegraphics[width=.48\textwidth]{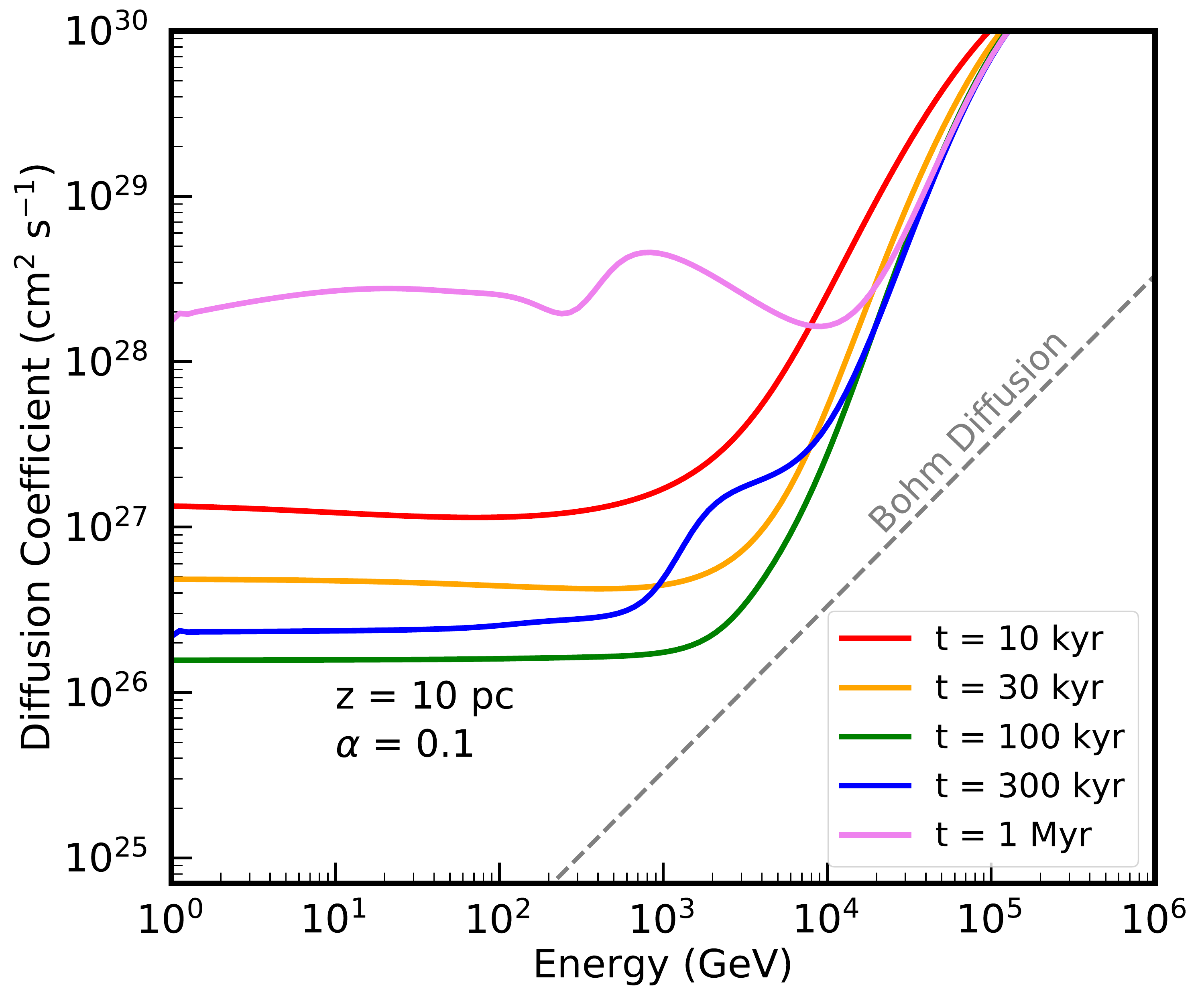}
\caption{The diffusion spectrum as a function of energy for different time snapshots for our default model with an injection efficiency of $\alpha = 0.1$. The grey dashed line indicates the Bohm limit. The diffusion coefficient is highly suppressed below $\sim$ 100 TeV and the suppression persists till 1 Myr. The diffusion coefficient rises rapidly for very high energy electrons above $\sim$ 10 TeV due to a combination of the exponential cutoff in the pulsar spectrum and very fast cooling of high-energy electrons. }
\label{fig:mom_dep}
\end{figure}

\begin{figure}[t]
\centering
\includegraphics[width=.48\textwidth]{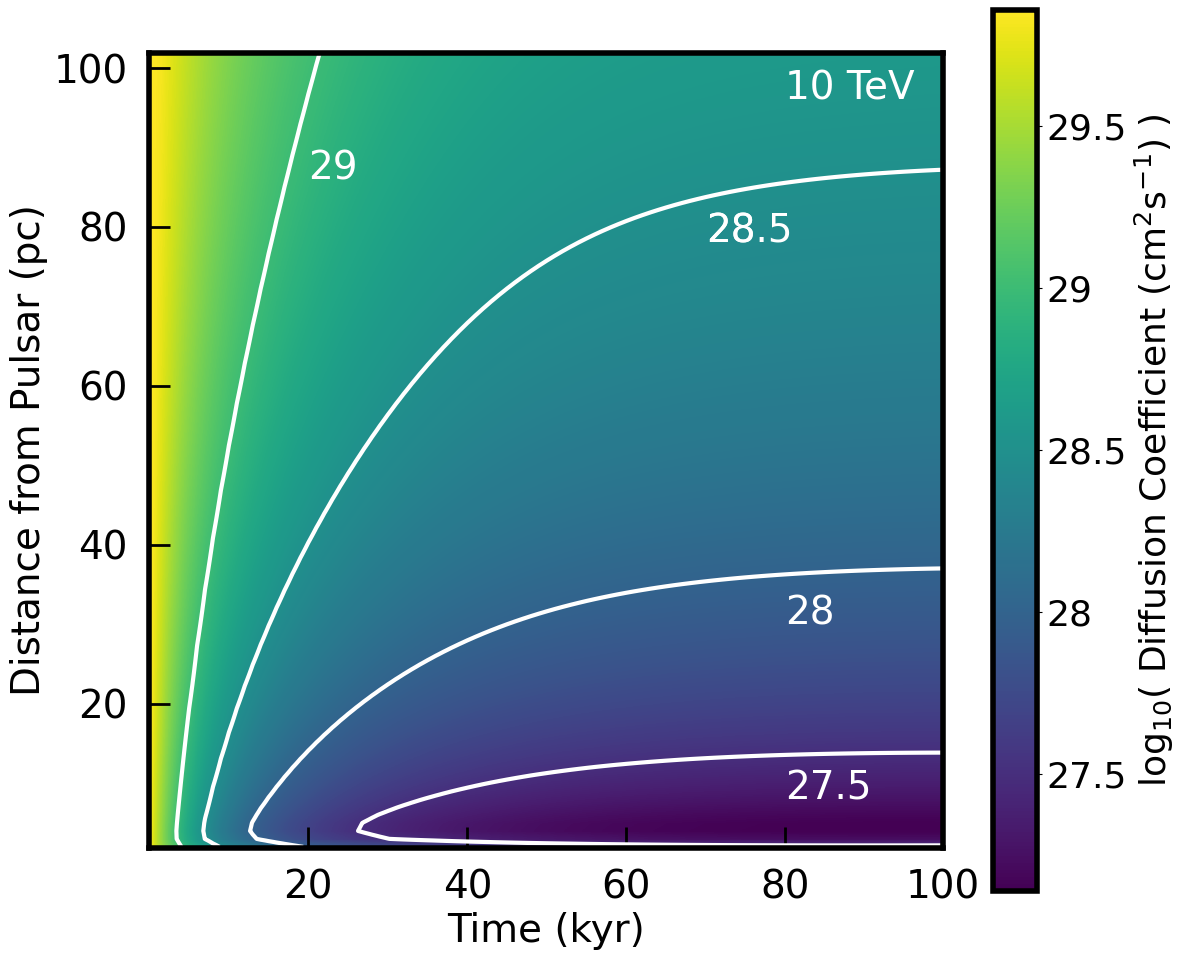}
\caption{A heatmap of the diffusion coefficient as a function of distance from the pulsar and age for our default 1D scenario with $\alpha = 0.1$. The diffusion coefficient is found to be suppressed by \mbox{$\sim$ 2--3} orders of magnitude within 10--20 pc of the pulsar. The resulting morphology of the halo is consistent with HAWC observations. However, the diffusion remains inhibited out to more than 100 pc from the pulsar, where the 1D approximation likely breaks down.}
\label{fig:heatmap_1D}
\end{figure}

In Fig. \ref{fig:eff_dep}, we show the evolution of the diffusion coefficient as a function of time at a constant distance of 10 pc, but for different source injection efficiencies. In our fiducial model, using a 10 \% efficiency, the diffusion coefficient for 10 TeV electrons is suppressed by more than two orders of magnitude at 340 kyr. For values of the injection efficiency as low as 3\%, we still obtain a significant (factor of $\sim$5) suppression in the diffusion coefficient, indicating that even relatively weak pulsars could significantly affect local cosmic-ray propagation. We note that these effects are non-linear. A model with a 10\% pulsar efficiency produces a decrease in local diffusion that is 50$\times$ larger than a model with a 1\% pulsar efficiency. This is due to the fact that the growth of turbulence is resonant. Larger cosmic-ray gradients lead to suppressed diffusion, which in turn causes more cosmic-rays to pile-up (leading to larger gradients). As a result, even small increases the pulsar power can significantly affect the resulting model.

For models with injection efficiencies exceeding $\sim$30\%, i.e $\alpha = 0.3$, the diffusion coefficient at 10~pc and 10 TeV falls below the Bohm limit. We note that the assumptions of resonant diffusion off of turbulent waves breaks down in such an environment. Thus, in our default models we set the minimum diffusion coefficient in any spatial and energy bin to be equal to the Bohm value at that energy. We note that even if the curve Fig~\ref{fig:eff_dep} does not show a value at the Bohm limit, the simulation may still be affected by this parameter choice if the Bohm limit is being reached in a radial bin closer to the pulsar.  

In Fig. \ref{fig:mom_dep}, we show the diffusion spectrum generated by our default model ($\alpha$~=~0.1) over the full energy range of our simulation. We immediately note several key results: (1) the diffusion coefficient is highly suppressed below energies of $\sim$100~TeV, (2) the minimum value of the diffusion coefficient is reached on timescales of $\sim$100~kyr across a wide-energy range, (3) the diffusion coefficient rises rapidly above $\sim$10~TeV, due to the significantly decreasing pulsar power (and highly efficient electron cooling) at higher energies. This result is potentially testable with upcoming HAWC and LHAASO observations, (4) the diffusion coefficient below $\sim$10~TeV is nearly energy-independent between 100-300~kyr, closely matching observations of energy-independent diffusion in the Geminga TeV halo that were first noted by~\cite{Hooper:2017gtd}. Notably, these spectral features may help differentiate cosmic-ray self confinement models from alternative explanations for inhibited diffusion within TeV halos.

In Fig. \ref{fig:heatmap_1D}, we show a two-dimensional representation of the diffusion coefficient at our standard energy of 10~TeV as a function of both time (after pulsar formation) and distance (from the pulsar) for our default simulation (with $\alpha$=0.1). At time $t$=0, the diffusion coefficient is high throughout the simulation volume, but it drops quickly in regions near the pulsar, achieving a minimum value of $\sim$10$^{27}$~cm$^2$s$^{-1}$ at an age of $\sim$100~kyr. The diffusion coefficient can continue to be significantly inhibited and much larger radii -- falling below a value of $\sim$10$^{28}$~cm$^2$s$^{-1}$ (70$\times$ lower than the standard ISM) out to a radius of $\sim$35~pc. Such a model matches (and even exceeds) the inhibition of diffusion observed near the Geminga and Monogem TeV halos~\cite{Abey:2017}, indicating the potential impact of cosmic-ray self-confinement in these simulations. 

\subsection{Robustness of Results}

Our 1D models indicate that the diffusion coefficient near pulsars: (1) can be suppressed by 2--3 orders of magnitude at critical energies near 10~TeV, (2) will remain suppressed on 1~Myr timescales consistent with the observation of TeV halos around mature pulsars, (3) can be suppressed even if only $\sim$10\% of the pulsar spindown power is converted into electrons. The surprising efficiency of cosmic-ray self-confinement indicates that these models are likely to remain robust even if less optimistic input parameters are considered. 

\begin{figure}[t]
\centering
\includegraphics[width=.48\textwidth]{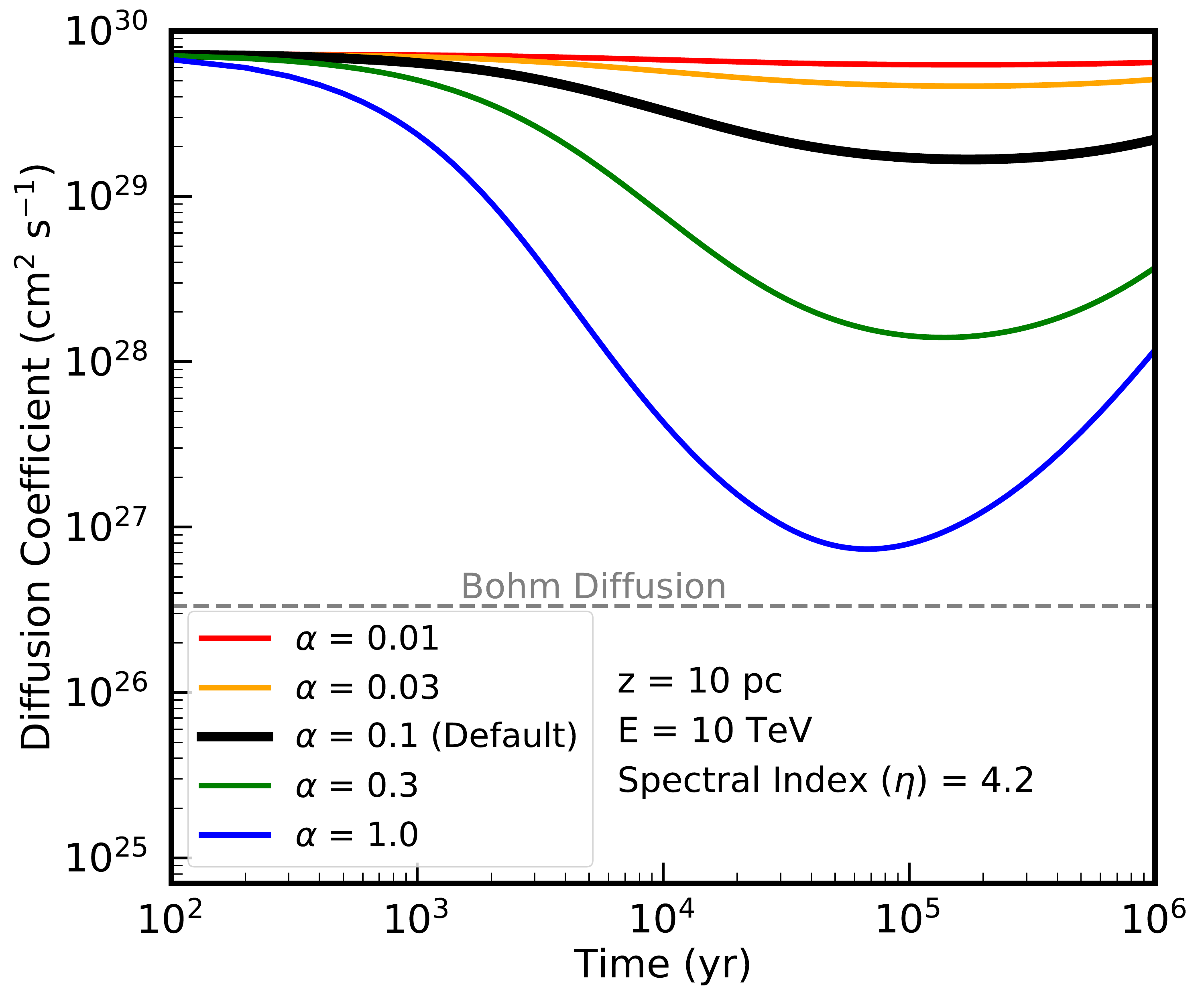}
\caption{The diffusion coefficient as a function of time at an energy of 10 TeV and a distance of 10 pc from the pulsar. The plot is shown for a softer pulsar spectral index, $\eta = 4.2$ than the one considered in Fig. \ref{fig:eff_dep}, which uses $\eta = 3.5$. We find that even with an extremely soft spectral index, the diffusion coefficient decreases by $\sim$ 2 orders of magnitude for a pulsar efficiency of 30\%, i.e $\alpha = 0.3$. The pulsar efficiency needed to inhibit the diffusion coefficient by 2 orders of magnitude is higher than our default model with $\eta = 3.5$ because the efficiency must increase to compensate for the softer spectrum.}
\label{fig:spectrum2.2}
\end{figure}

\begin{figure}[t]
\centering
\includegraphics[width=.48\textwidth]{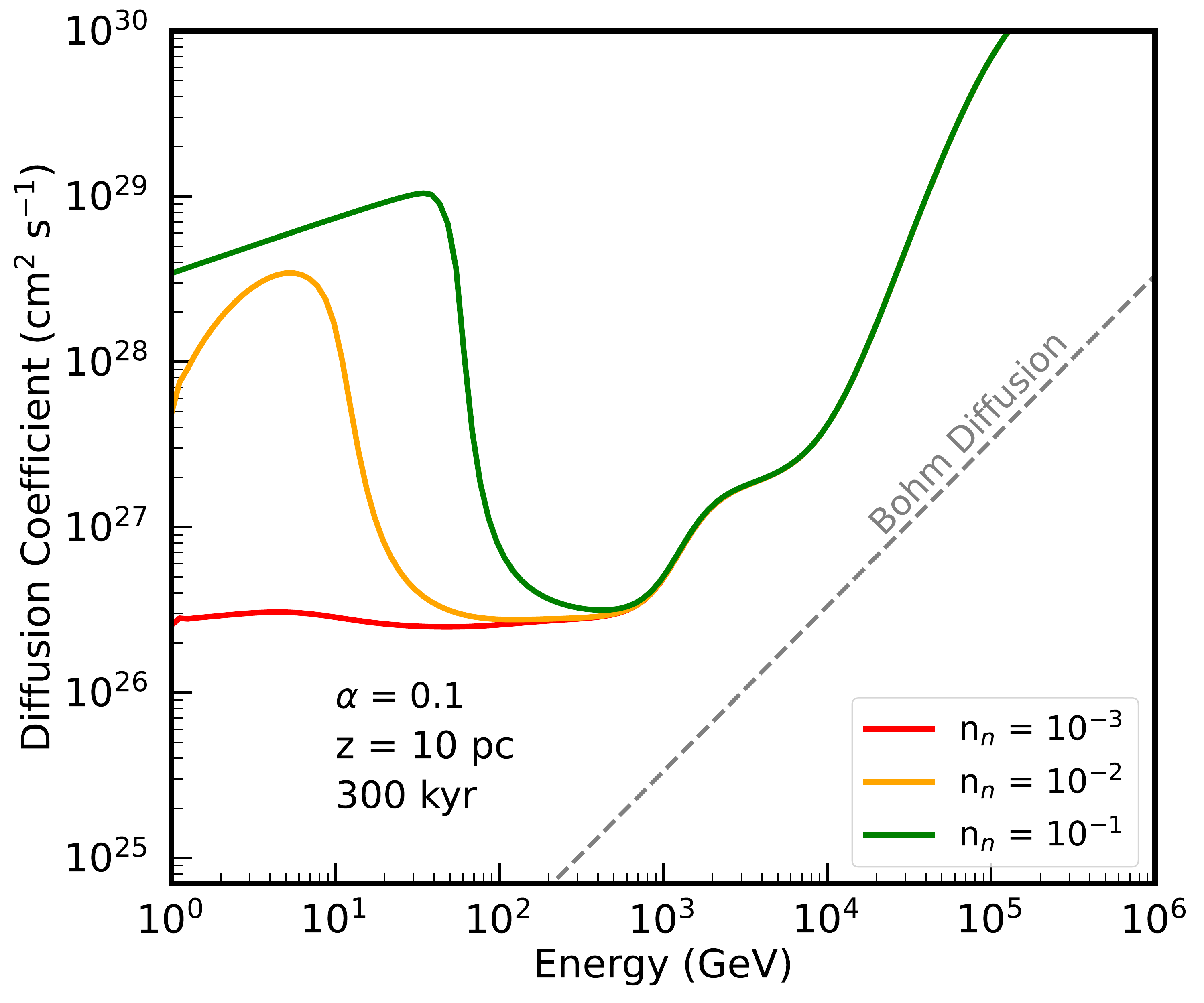}
\caption{The diffusion coefficient as a function of energy in a model where we include IND damping. The result is shown at 300 kyr and 10~pc from a pulsar with an efficiency of $\alpha$=0.1. The ion number density, $n_i$ is set to 1 cm$^{-3}$. We show results for neutral atom number densities, $n_n$ = 0.1, 10$^{-2}$ and 10$^{-3}$ cm$^{-3}$ (corresponding to a 10\%, 1\% and 0.1 \% neutral fraction). For a 0.1\% neutral fraction, IND damping has a negligible impact on our model and the diffusion spectrum at 300 kyr very similar to Fig. \ref{fig:mom_dep}. When the neutral fraction increases to 1\%, the low energy ($\lesssim 1$ TeV) flux begins to be significantly affected. However, the diffusion coefficient at 10 TeV is not impacted for any of the neutral fractions considered.}
\label{fig:IND_damping}
\end{figure}

In Figure~\ref{fig:spectrum2.2} we show the diffusion coefficient as a function of time in a model where the pulsar electron injection spectrum has a momentum index p$^{-4.2}$, which is significantly softer than our default value of p$^{-3.5}$. Our default value is more consistent with observations of the Geminga and Monogem TeV halos~\cite{Hooper:2017gtd}, while a softer value may be motivated by GeV observations of Geminga~\cite{DiMauro:2019yvh} as well as the spectrum of diffuse emission from the TeV halo population~\cite{Linden:2017blp}. However, we stress that our choice of a p$^{-4.2}$ momentum spectrum is even more pessimistic than advocated by the results of~\cite{DiMauro:2019yvh, Linden:2017blp}, as we assume that the soft cosmic-ray injection spectrum continues all the way to 1~GeV, while previous models have utilized a broken power-law to decrease the electron power below energies of $\sim$500~GeV. Even in this extremely pessimistic case, we find that the diffusion coefficient at 10 TeV can be inhibited by 2--3 orders of magnitude so long as the electron efficiency is increased to offset the softened spectrum.

In Figure~\ref{fig:IND_damping}, we show the diffusion spectrum for models that additionally include ion-neutral damping (IND). The total damping term is taken as the sum of the IND and NLD terms, as defined in Section~\ref{sec:1D_transport}. We find that the IND term generally has no effect at TeV energies, but can considerably affect diffusion in the GeV range. This result is intriguing, as it indicates that TeV halos in regions with low neutral gas densities (such as Geminga) may also produce GeV halos~\cite{DiMauro:2019yvh}, while TeV halos in dense regions of the ISM may lack GeV counterparts. This observation appears unique to cosmic-ray self-confinement models and could be tested by future Fermi-LAT data.

\begin{figure}[t]
\centering
\includegraphics[width=.48\textwidth]{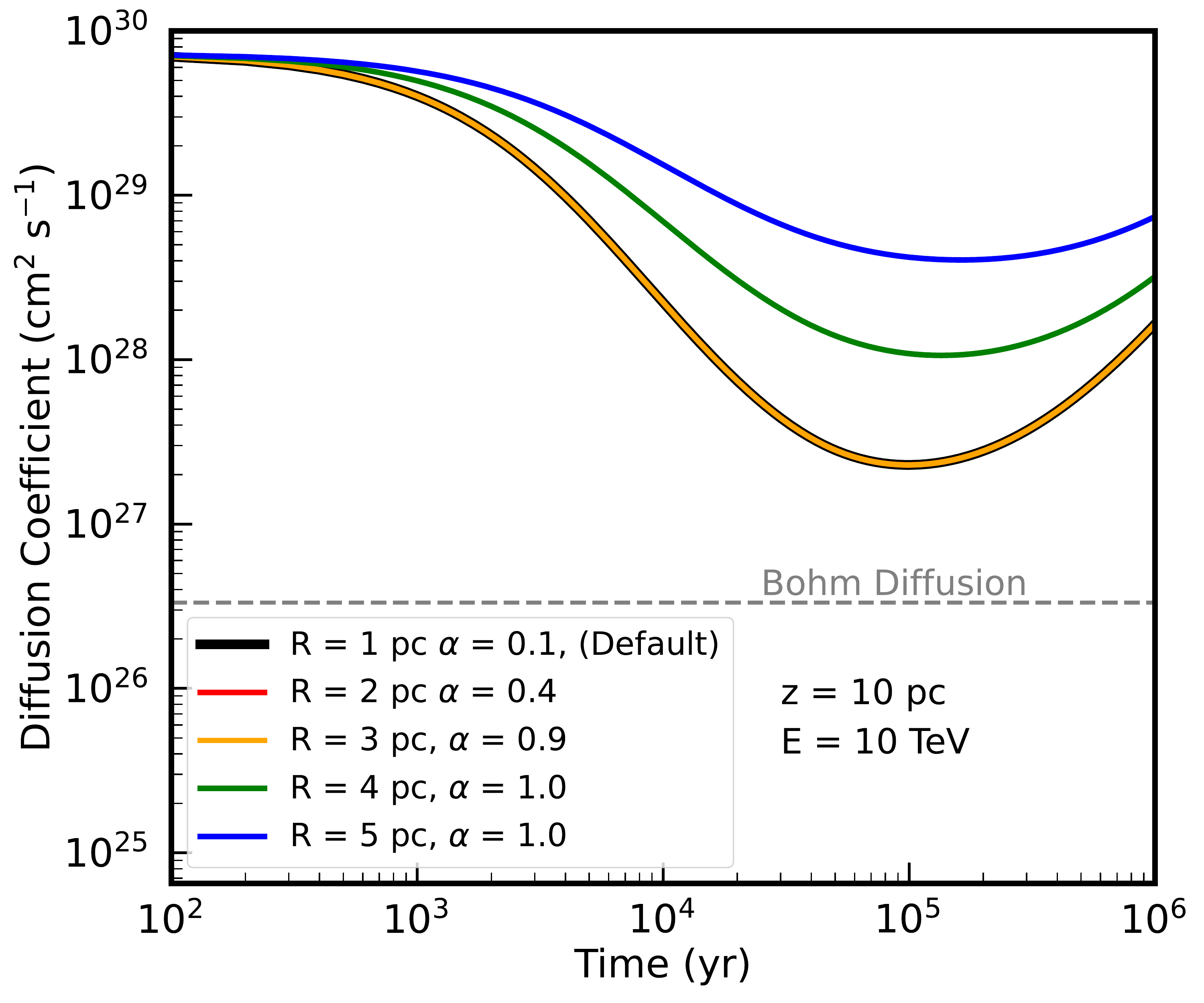}
\caption{The diffusion coefficient as a function of time for different flux tube radii $R$ and source injection fractions $\alpha$. Increasing the flux tube radius lowers the suppression of the diffusion coefficient by increasing the propagation volume. This can initially be compensated for by increasing the injection fraction $\alpha$. For example, the black and yellow solid lines correspond to flux tube radii of 1 pc and 3 pc respectively, but have an identical diffusion coefficient because the increasing radius is compensated by increasing $\alpha$ from 0.1 to 0.9.}
\label{fig:tube_rad}
\end{figure}

Finally, in Figure~\ref{fig:tube_rad} we investigate models with significantly wider flux tube radii. Within a 1D model, such a change is degenerate with changing the efficiency ($\alpha$) of the pulsar, because the electrons are evenly distributed throughout the radius of the flux tube. However, this degeneracy breaks down when the pulsar efficiency approaches unity. Further increases in the flux tube radius can not be compensated by changing the pulsar energetics, and the degree of cosmic-ray self-confinement begins to decrease. However, even for flux tubes with a radius of 5~pc, we can obtain a diffusion coefficient that is suppressed by more than an order of magnitude at a distance 10~pc. Such a model provides a hint regarding the transition from 1D to 3D models -- indicating that in regimes where the flux tube approximation begins to break down, the pulsar may continue to significantly inhibit local diffusion.

\begin{figure}[t]
\centering
\includegraphics[width=.48\textwidth]{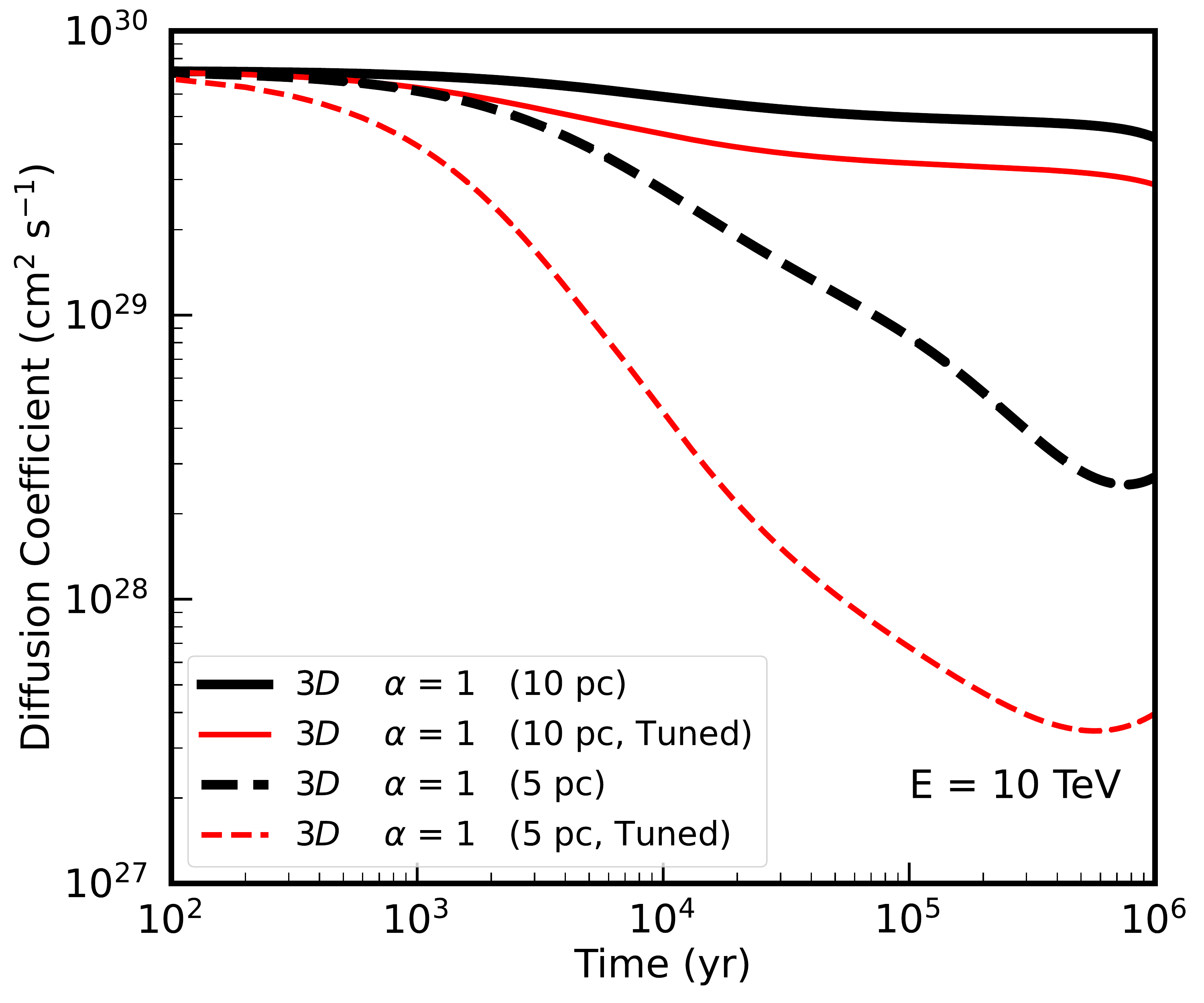}
\caption{Evolution of the diffusion coefficient as a function of the pulsar age for 3D diffusion models. We assume a 100\% pulsar injection efficiency ($\alpha = 1.0$). Results are shown at 5 pc and 10 pc for two models: the fiducial 3D model and a `tuned' model adjusted to have extreme parameters. The fiducial model has $B = 1\mu G$ and a spectral index $\eta = 3.5$. The tuned model has a spectral index of $\eta = 3.0$, a magnetic field of $B = 0.5 \mathrm{\mu G}$ and a minimum pulsar power of 10~GeV. The 3D model produces a very low suppression (less than a factor of 3) of the diffusion coefficient at 10 TeV for 10 pc distance. However, this model is capable of producing very inhibited diffusion coefficients within the inner $\sim$ 5 pc. }
\label{fig:3D_extreme}
\end{figure}

In view of the results above, we conclude that, within the context of 1D simulations, the self-generation mechanism can be highly efficient in suppressing the diffusion coefficient surrounding TeV halos by 2--3 orders of magnitude at distances between 10--20~pc from the pulsar. Moreover, the diffusion coefficient is maximally suppressed on timescales of $\sim$100~kyr, consistent with the observations of extremely suppressed ISM diffusion in the TeV halos that surround multiple middle-aged pulsars~\cite{2017PhRvD..96j3016L, 2017ATel10941....1R, 2018ATel12013....1B, 2019MNRAS.488.4074F, 2019PhRvD.100d3016S, Fang:2021qon, LHAASO:2021crt}. We also note that diffusion remains relatively efficiency within the first $\sim$10~kyr, which may explain the lack of an observed TeV halo around the Crab nebula and the observation of only a dim halo around Vela (although we note that the PWN, which is not modeled in this analysis, also plays a significant role in electron propagation at such an early stage).

\subsection{3D model}
\label{sec:result_3D}

We note that in 1D models (where the volume increases linearly with distance), the diffusion coefficient remains inhibited more than 100~pc from the pulsar. Such a scenario appears unphysical, as the correlation length of galactic magnetic field turbulence likely falls below 100~pc, even in particularly low-density regions of the galaxy~\cite{Iacobelli:2013fqa,Haverkorn:2008tb}. Additionally, the spherically symmetric nature of several TeV halo observations motivates the investigation of 3D models. 

In Figure~\ref{fig:3D_extreme}, we show the evolution of the diffusion coefficient as a function of pulsar age for our 3D case with two different model parameters. One is the 3D fiducial parameters outlined in \ref{sec:numerics} where the pulsar injection efficiency, $\alpha = 1.0$, $B = 1 ~ \mu G, \sigma = 1$ pc, $\tau = 10$ kyr and the background ISM diffusion coefficient is $D = 3.466 \times 10^{28} p^{1/3}$ cm$^2$ s$^{-1}$. The momemtum dependence of the pulsar injection for this fiducial model is $\propto p^{-3.5} \exp{(- \frac{p}{100 ~ \mathrm{TeV}})}$. Additionally, we consider a ``tuned" model where we adjust several parameters to optimistic values that give the maximum suppression in the diffusion coefficient, namely $B = 0.5 \mathrm{\mu G}$, $\tau = 5$ kyr, a minimum injection momentum $p_{min} = 10$ GeV and a momentum dependence of the pulsar injection $\propto p^{-3.0} \exp{(- \frac{p}{100 ~ \mathrm{TeV}})}$. For both the cases, we plot the diffusion coefficient as a function of time at both 5 pc and 10 pc from the pulsar. We note that this plot is similar to Figure~\ref{fig:eff_dep} in the 1D case. 

\begin{figure}[t]
\centering
\includegraphics[width=.48\textwidth]{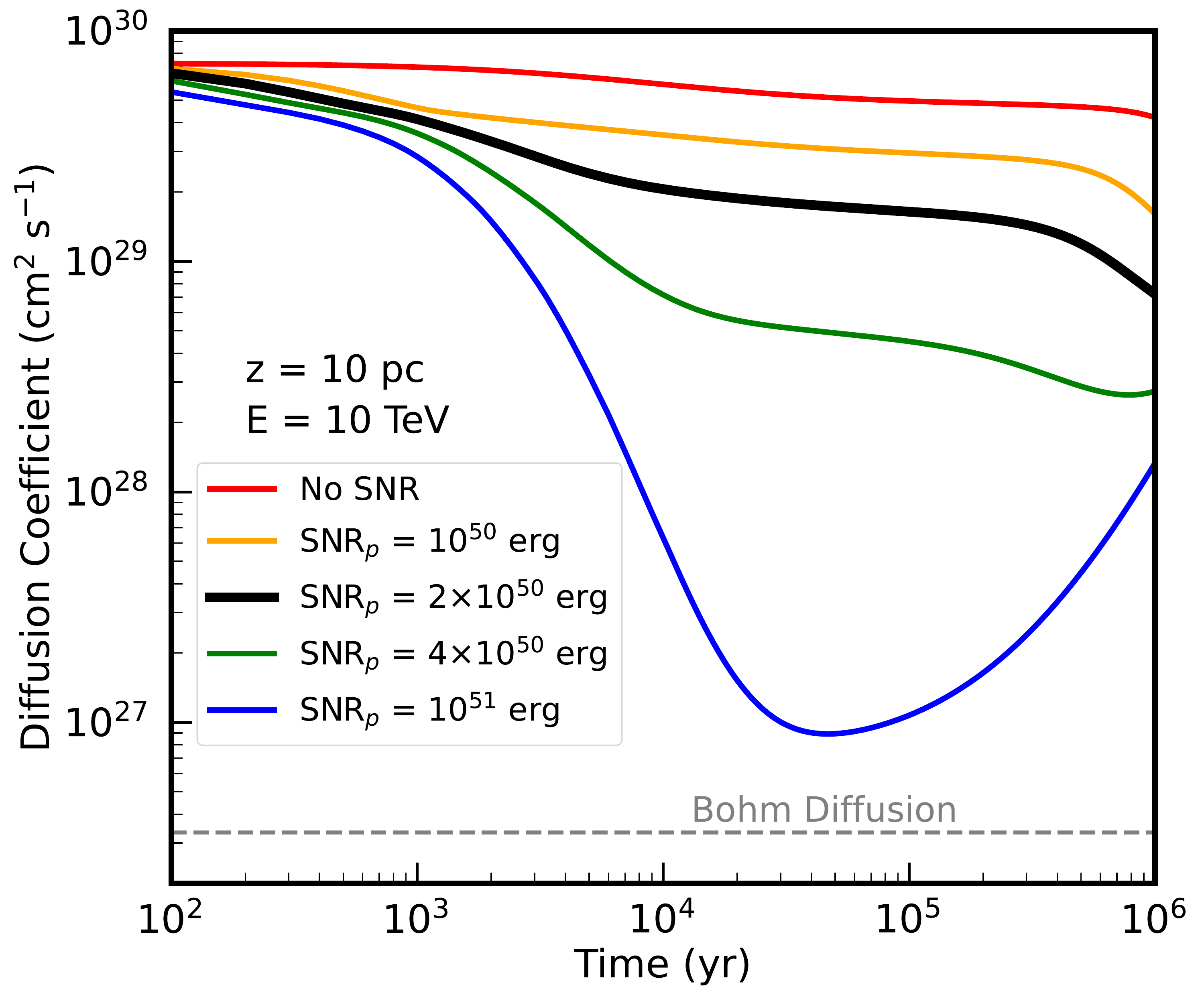}
\caption{The diffusino coefficient as a function of pulsar age at 10 pc and 10 TeV for the case of 3D propagation with SNR proton injection. Results for different SNR injection power into protons (denoted by SNR$_p$) are shown. We set $B = 1 \mu G$ and the SNR spectral index, $\eta_{SNR} = 4.2$. In order to suppress diffusion at 10 pc by more than one order of magnitude, the SNR must convert 4 $\times 10^{50}$ erg (or 40\% of the SNR kinetic energy) into protons, which exceeds standard SNR models.}
\label{fig:SNR_eff}
\end{figure}

Focusing on the black solid and dashed plots that correspond to the fiducial model, we find that the diffusion coefficient at 10 pc is suppressed by less than a factor of $2$ over the 1~Myr timescale of our simulation. However, at 5~pc, the diffusion coefficient is suppressed by more than 1.5 orders of magnitude until 1 Myr. We find that even our extremely ``tuned" model does not significantly inhibit cosmic-ray diffusion at 10~pc (decreasing the diffusion coefficient by only a factor of $\sim$3). However, it drastically decreases the diffusion coefficient (by more than two orders of magnitude) at 5~pc. Thus, our simulations indicate that in 3D models, the pulsar can significantly inhibit cosmic-ray diffusion on $\sim$5~pc scales, but is insufficiently powerful to inhibit diffusion on the 10--20~pc scales that are consistent with TeV halo observations~\cite{Abey:2017}.

\subsection{3D model with Supernova Remnant}
\label{sec:result_3D&SNR}

Because cosmic-rays from the pulsar appears to be insufficient to significantly decrease the diffusion coefficient on 10--20~pc scales, we consider the possibility that protons from the coincident supernova may additionally contribute to lowering the diffusion coefficient through a similar process. Utilizing an identical pulsar model as described for the `default" 3D model ($\alpha$ = 1.0), we add an SNR with a total kinetic energy of 10$^{51}$ erg. The momentum dependence of the proton injection spectrum by the SNR is taken to be $\propto p^{-4.2} \exp(-\frac{p}{10^7 ~\mathrm{GeV}})$. Protons from the SNR are treated identically as the electrons in our default model, with the exception that they do not cool (because hadronic losses are inefficient on these time- and density scales). 

In Figure~\ref{fig:SNR_eff}, we show the results for this model at our default radius of 10~pc and energy of 10~TeV for different assumptions regarding the fraction of the SNR kinetic energy that is converted into TeV protons. In all cases, we find that the SNR significantly increases the suppression of the local diffusion coefficient. However, a relatively high conversion (40\% of the total SNR) power must be converted into cosmic-ray protons to produce the factor of 30 suppression in the diffusion coefficient that begin to make these models consistent with TeV halo observations. The full two-order of magnitude suppression of the diffusion coefficient is only possible if the entire kinetic energy of the SNR is converted into protons.

\begin{figure}[t]
\centering
\includegraphics[width=.48\textwidth]{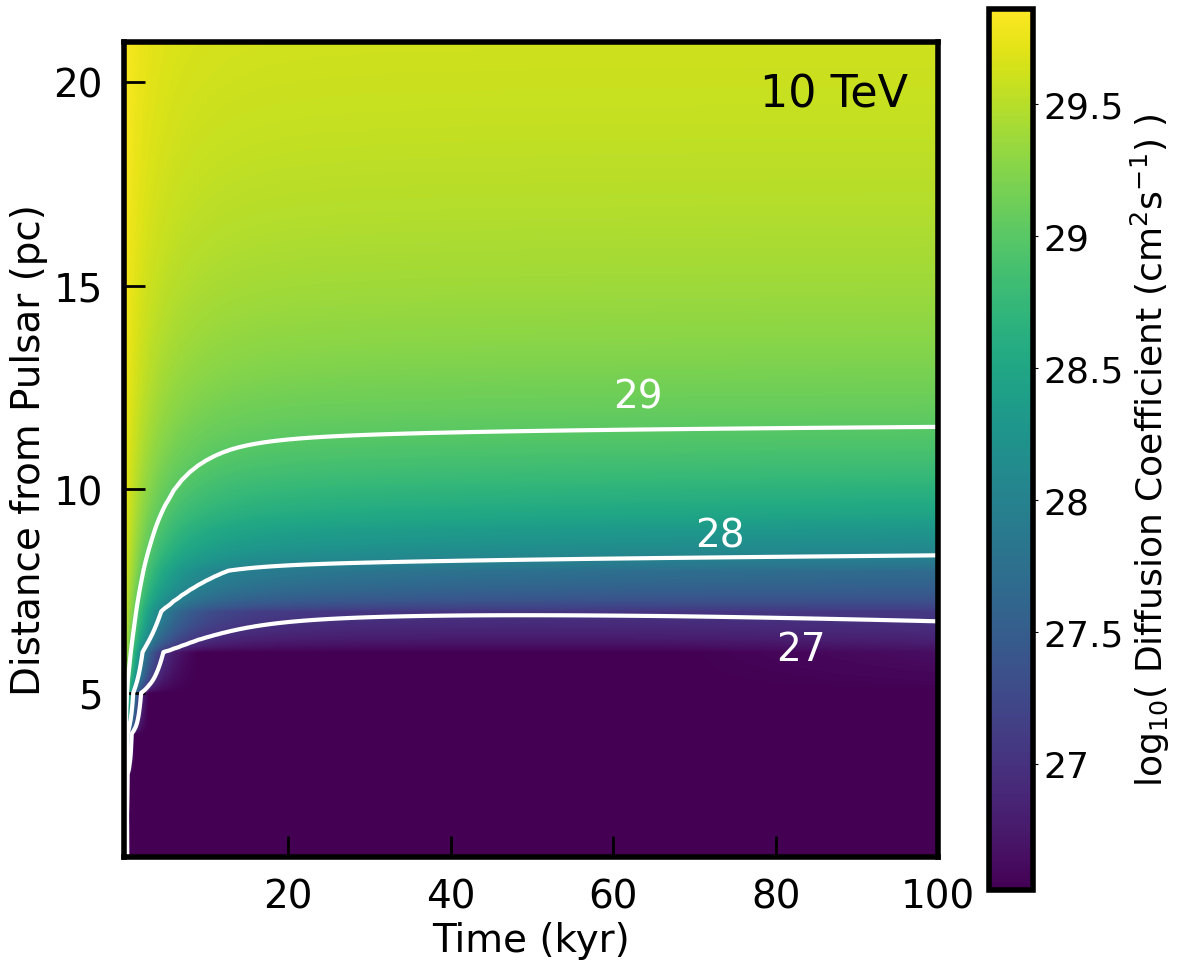}
\caption{Colormap for $D$ as a function of distance from the pulsar and age. This plot assumes an optimistic $\xi_{CR} = 0.4$. The diffusion coefficient can be highly suppressed in the inner $\sim$ 5 pc. The suppression is about an order of magnitude at $\sim$ 10 pc. The diffusion coefficient quickly rises back to its ISM values after 10 pc.}
\label{fig:heatmap_3D}
\end{figure}

Finally, in Figure~\ref{fig:heatmap_3D}, we show a two-dimensional heatmap of the diffusion coefficient as a function of time and distance from the pulsar in our 3D model with a supernova contribution. We set the conversion efficiency of proton kinetic energy into protons to $\xi_{p} = 0.4$. These results demonstrate two key (and intuitive) differences between our 1D and 3D models: (1) in 3D models that include a significant SNR components, the diffusion coefficient near the source becomes extremely small, saturating the Bohm limit for regions that are within 5~pc of the pulsar on timescales exceeding 100~kyr. In addition to demonstrating that SNR activity can significantly affect local diffusion - this indicates that our simulation may not be robust, as it is difficult to utilize our diffusive model under such extreme conditions, and (2) unlike the case of 1D diffusion, the quickly expanding volume of 3D diffusion localizes the effect of cosmic-ray induced self-confinement to a region within $\sim$10~pc of the central source. The diffusion coefficient rapidly returns to standard ISM values by $\sim$20~pc from the pulsar. 

\section{Discussion and Conclusions}

In this paper, we have re-examined models of cosmic-ray induced self-confinement. We have corrected an error in the previous study of Ref.~\cite{Evoli:2018aza}, which utilized an unphysically large non-linear damping term. The net effect of this change demonstrates that pulsars are significantly more capable of inhibiting cosmic-ray diffusion by more than a factor of 100, consistent with TeV halo observations. Moreover, once diffusion is suppressed by pulsar activity, it remains inhibited on nearly Myr timescales, consistent with the observation of TeV halos around mature pulsars. These two findings paint a much more optimistic picture regarding the potential for pulsars to confine their own cosmic-ray population, compared with previous work.

Excitingly, our models indicate several testable features that are unique to the self-confinement scenario. These may be used to distinguish self-confinement models from alternative models with pre-existing low diffusion coefficients \cite{Abeysekara:2017hyn,Tang:2018wyr,Giacinti:2018yjc,Lopez-Coto:2017pbk}, anisotropic diffusion models \cite{Liu:2019zyj} and models incorporating the transition from rectilinear to diffusive propagation \cite{2021:Recchia}. First, our model predicts that in the absence of IND damping, the diffusion coefficient will be roughly energy independent for $p \lesssim 1$ TeV, and the diffusion coefficient will rapidly rise near the exponential energy cutoff of the pulsar (Fig. \ref{fig:mom_dep}). Second, the presence of stronger IND damping in dense regions of the ISM will prevent the observation of GeV emission from TeV halos (Fig. \ref{fig:IND_damping}) --- producing a bifurcation of TeV halo observations that depends on the local gas density. Third, our models indicate that standard TeV halos should not form around systems much younger than $\sim$10~kyr. All of the above predictions are exclusive to this mechanism and are potentially testable with future TeV observations.

\subsection{Comparison of 1D and 3D Models}
While our results indicate that cosmic-ray self confinement models are robust to changes in the initial pulsar spectrum, the presence of IND damping, and the width of the 1D flux tube radius, our results strongly depend on whether particle diffusion is initially confined to 1D flux tubes, or initially diffuses efficiently in all three spatial directions.

In 1D models, many details of the self-generation mechanism are consistent with TeV halo observations: (1) the diffusion coefficient is suppressed by 2--3 orders of magnitude, (2) the diffusion coefficient reaches a minimum value at $\sim$100~kyr after pulsar formation. Moreover, we note that our 1D framework is a reasonable approximation of reality, as propagation through 1D flux tubes likely dominates diffusion on distances less than the 1--100~pc coherence length of the galactic magnetic field~\cite{Lopez-Coto:2017pbk,Iacobelli:2013fqa,Haverkorn:2008tb}. On the other hand, these models may appear insconsistent with data, because observations indicate that TeV halos are roughly spherical sources.

However, we \emph{stress} that our 1D diffusion models are very different (and more consistent with data) than competing flux-tube models~(e.g.,~\cite{Liu:2019zyj}) that do not use cosmic-ray self-confinement. In previous models, the flux tube must be oriented directly towards Earth (with an offset less than $\sim$5$^\circ$), to ensure that the halo does not extend too far in the transverse (observable) direction. Such models are difficult to rectify with observations which indicate that many pulsars produce observable TeV halos. 

In our 1D models, the preferential direction for particle diffusion is irrelevant. The cosmic-ray self-confinement mechanism efficiently inhibits cosmic-ray propagation along this direction until it is \emph{also} inhibited. Notably, the degree of inhibited diffusion (by a factor of $\sim$100) is similar to the predicted efficiency of diffusion in directions perpendicular standard 1D flux tubes~\cite{Liu:2019zyj}. The sum of these effects produce a cosmic-ray population whose diffusion is inhibited in every direction. Therefore, it is \emph{critical} to note that the $\gamma$-ray emission morphology from our 1D models is not yet known. It could potentially even be spherical, if cosmic-ray self-confinement forces the diffusion coefficients in every direction towards similar values. The interaction of cosmic-ray self-confinement mechanisms with a multi-dimensional diffusion tensor, and the comparison of such a model with observed TeV halos, is important work which we leave for future publications.

Finally, while 1D models predict that diffusion continues to be inhibited on distance scales that are much larger than the observed size of TeV halos, this result is likely consistent with data for two reasons. First, the edges of TeV halos are not well-defined and may be significantly affected by the flux sensitivity of each $\gamma$-ray instrument. Indeed subsequent HAWC observations of Geminga and Monogem found significantly larger TeV halos, which was consistent with the improved exposure of each source~\cite{Abey:2017}. Second, we expect the flux tube approximation to break down on larger distance scales, an effect which will quickly terminate the region where diffusion is inhibited, as demonstrated from our 3D models. 

In 3D models, on the other hand, the pulsar is not energetic enough to significantly inhibit cosmic-ray diffusion -- even when the model is strongly tuned to maximize the inhibition of diffusion near 10~TeV. To produce reasonable models, we added a contribution from the coincident SNR, which adds an additional 4$\times$10$^{50}$~ergs of cosmic-ray energy into the local ISM (for our best-fit $\xi_{CR} = 0.4$ value). In this case, we find that diffusion is significantly inhibited on $\sim$5~pc scales, reaching the Bohm limit. However, the large volume of 3D simulations still prevents our model from significantly affecting cosmic-ray diffusion on 20~pc scales. 

We note, however, that our 3D model is \emph{maximally} pessimistic for cosmic-ray self-confinement models. The simulation begins with a diffusion coefficient that is unsuppressed in every direction around the central source, allowing cosmic-rays to quickly be diluted as they move outword in every direction. Models that include inhibited diffusion in one direction (even on small distance scales), will force the cosmic-ray flux to build up, producing larger gradients (and thus significantly inhibited diffusion) throughout the simulation volume. Notably, these effects can be strongly non-linear, as shown by the significant changes in the diffusion coefficient which result from relatively small changes in the pulsar efficiency. Finally, we note that these models do not include SNR shocks or other molecular dynamics that may also significantly increase magnetic turbulence and inhibit cosmic-ray diffusion.

To conclude, our 3D models where the self-confinement scenario is ineffective at inhibiting cosmic-ray diffusion on 10--20~pc scales are models where the initial diffusion coefficient is given by the standard ISM value in every spatial direction surrounding the pulsar source. However, any adjustments to this model, including diffusion coefficients that are initially suppressed in one or more spatial directions, or contributions from SNR/PWN activity that lower the initial diffusion coefficient near the pulsar source -- will unambiguously act to strengthen the self-confinement mechanism by allowing the particle density to build up in compact regions. Thus, we conclude that in realistic models, the cosmic-ray self-confinement mechanism is capable of driving the inhibited diffusion coefficients observed out to 10--20~pc from pulsar sources.

\subsection{Future Improvements}
Finally, we note a number of potential improvements which lie beyond the scope of this model, but which may be considered in upcoming work. First, we note that some pulsars (including Geminga), have large proper velocities which have significantly removed them from their natal SNR by the $\sim$300~kyr observation time of their TeV halos. Such models make the inclusion of turbulence from the initial SNR less realistic -- though the SNR may stil act as important seed to drive down the initial diffusion coefficient. Future models must consider the impact of pulsar proper motion through the ISM. Secondly, in our models, we have not yet considered the potential interplay between 1D and 3D diffusion on the boundary between different flux tubes. However, our results (which indicate that pulsars produce diffusion that is too inhibited in 1D simulations but not inhibited enough in 3D simulations), strongly motivate such an extension. 

\appendix

\section*{Acknowledgements}
We would like to thank Felix Aharonian, Elena Amato, John Beacom, Pasquale Blasi, Ilias Cholis, Dan Hooper, Alexandre Marcowith, Pierrick Martin, Sarah Recchia and Takahiro Sudoh for helpful comments which greatly improved the quality of this manuscript. We would especially like to thank Carmelo Evoli and Giovanni Morlino for a number of comments, code comparison tests, and discussions that motivated and significantly improved the quality of this manuscript. TL is partially supported by the Swedish Research Council under contract 2019-05135, the Swedish National Space Agency under contract 117/19 and the European Research Council under grant 742104.  This project used computing resources from the Swedish National Infrastructure for Computing (SNIC) under project Nos. 2020/5-463 and 2021-1-24 partially funded by the Swedish Research Council through grant no. 2018-05973.

\newpage
\bibliography{main_prd.bib}

\end{document}